\documentclass{Interspeech}
\usepackage{graphicx}
\usepackage{multirow}
\usepackage{array}

\interspeechcameraready

\title{Stepback: Enhanced Disentanglement for Voice Conversion via Multi-Task Learning}

\author[affiliation={1}]{Qian}{Yang}
\author[affiliation={2}]{Calbert}{Graham}

\affiliation{}{Shanghai International Studies University}{China}
\affiliation{Phonetics Laboratory}{University of Cambridge}{UK}

\email{0211110032@shisu.edu.cn, crg29@cam.ac.uk}
\keywords{Voice Conversion, Variational Autoencoder, Content Disentanglement, Multi-Task Learning}

\usepackage{comment}

\begin{document}

\maketitle

\begin{abstract}
    
    Voice conversion (VC) modifies voice characteristics while preserving linguistic content. This paper presents the Stepback network, a novel model for converting speaker identity using non-parallel data. Unlike traditional VC methods that rely on parallel data, our approach leverages deep learning techniques to enhance disentanglement completion and linguistic content preservation.
    
    The Stepback network incorporates a dual flow of different domain data inputs and uses constraints with self-destructive amendments to optimize the content encoder. Extensive experiments show that our model significantly improves VC performance, reducing training costs while achieving high-quality voice conversion. The Stepback network's design offers a promising solution for advanced voice conversion tasks.

\end{abstract}

\section{Introduction}

Voice conversion (VC) refers to the technology of altering the voice characteristics of one speech utterance to match another, without changing the linguistic content. This technology is widely used in various fields, including speaker disguise, singing voice conversion, computer-assisted pronunciation training, and voice cloning. VC is considered a core task alongside Automatic Speech Recognition (ASR) and Text-to-Speech (TTS) synthesis. VC models benefit significantly from advances in these areas. Substantial progress and impressive performance have been achieved in VC, particularly in the conversion of emotion, timbre, and style\cite{17}\cite{8}. In this paper, we focus on the conversion of speaker identity, specifically on content disentanglement.

Voice conversion based on parallel data of paired speakers has been extensively studied. However, due to the constraints of frame-wise mapping and limitations in training data, much more attention has shifted towards VC with non-parallel data. Non-parallel data can have varying sequence lengths for the source and the target. This shift is especially notable in those methods leveraging deep learning. For instance, Generative Adversarial Networks (GANs) play a significant role in voice conversion by enabling the generation of matching and realistic high-quality speech waveforms. The generator network in a GAN can be trained to convert source speech into target speech, while the discriminator ensures the converted speech is indistinguishable from real speech of the target speaker. However, GANs are known for their training difficulties and challenges in maintaining stability\cite{5}\cite{6}.

Compared to this direct transformation methodology, Variational Autoencoders (VAEs) offer advantages in their suitability for any-to-any voice conversion and their generalizability to unseen speakers. VAEs consist of two main parts: a content encoder and a decoder. The content encoder processes source speech, transforms it into a latent representation, and removes speaker information. The decoder takes the speaker identity, combines it with the latent representation, and reconstructs the speech\cite{9}. A notable VAE approach is disentangling speaker and content representations using instance normalization, which supports one-shot learning and performs well on unseen data\cite{18}.

As a significant milestone, AutoVC combines the matching accuracy of GANs with the training ease of VAEs. It achieves this by deliberately designing a bottleneck, which enables zero-shot style transfer\cite{4}. AutoVC excels in conditioning speaker embeddings. Moreover, voice conversion has recently been considered from a broader perspective. One attractive method is the Blow technique\cite{10}, which normalizes flows of non-parallel and many-to-many data. Large models such as Diff-VC\cite{11}, a diffusion-based voice conversion model, and its variants outperform many existing models in terms of naturalness and similarity. However, these models are highly complex, with numerous parameters to tune, making them challenging to train. Additionally, there are models focused on achieving low latency. These models are more applicable in real-world scenarios and achieve comparable results in certain evaluation metrics\cite{yang2024streamvcrealtimelowlatencyvoice}.

Broadly speaking, most of these models are implemented under the concept of multi-task learning\cite{13}. Multi-task learning is a robust method for improving model training, reducing the risk of overfitting, and saving training costs across multiple tasks\cite{kendall2018multitasklearningusinguncertainty}. Typically, encoders are associated with speech recognition, while decoders are linked to text-to-speech models\cite{16}. This approach performs well when dealing with a single goal, such as many-to-one synthesis. More subtly, adversarial training is quite common in various VC models, including GANs and VAEs. This technique combines losses from several micro-models and aims to maximize one of the losses to benefit the main task. It helps focus training attention and avoids oversmoothing\cite{3}.

In the subsequent stage, we incorporate two data sources: one with the same speaker ID as the source speech and the other with a random speaker ID. We allow two identical decoders to operate simultaneously, taking latent representations from the content encoder. The first speaker ID is processed by the first decoder. After decoding, we calculate the distance between the converted speech and the source speech, denoted as Loss A. The second speaker ID is processed to generate a newly constructed speech. This speech is then classified by a pretrained classifier, using cross-entropy as Loss B.

Our overall goal is divided into two parts:
\begin{enumerate}
    \item Train the content encoder and decoder to minimize Loss A between source and target. This training makes the encoder and decoder proficient in handling speaker information and reconstructing speech.
    \item Train the content encoder and decoder to maximize Loss A while minimizing Loss B. This dual objective enables the system to eliminate residual speaker traces as much as possible while maintaining functionality.
\end{enumerate}

To guide our investigation and address the relevant challenges in voice conversion, we propose the following research questions:
\begin{enumerate}
    \item Does the classifier, which is typically added to be fooled by the latent representation from the content encoder, result in performance damage, impurities, and oversmoothing?
    \item How does the proposed dual flow of different domain data inputs address the problem identified in Question 1? The dual flow, which is the two data sources mentioned before, is processed by the same decoder and undergoes allied multi-task learning with two combined loss functions. One loss function employs relaxed adversarial learning that separates the original fierce adversarial interaction. The other uses strengthened positive learning.
\end{enumerate}

\section{Proposed method}

\subsection{Preparatory Stage}

For simplicity, we denote the content encoder input (source speech), the speaker identity, and the converted speech with x, i, y $\epsilon$ X, I, Y, respectively. Additionally, (x, i) $\epsilon$ T, the training set, and y $\epsilon$ Y, the output set.

\subsubsection{Stabilizing Encoder and Decoder}

In the preparatory stage, we use x and i, both from the same speaker, to train the content encoder and decoder. This stage follows the classic design of the Variational Autoencoder (VAE), excluding the classifier used in Chou's stage 1\cite{1}. The reason for excluding this latent representation classifier will be discussed in Section 2.1.2. The objective of the content encoder in this stage is to convert the sequence x into a latent representation, while the decoder's role is to reconstruct the speech from the latent representation produced by the content encoder and the speaker identity input. This is achieved by minimizing the reconstruction loss, as shown below (we minimize the Mean Absolute Error as suggested), where $\theta_{enc}$ and $\theta_{dec}$  indicate the parameters of the encoder and decoder respectively: 

\begin{equation}
L_{pre}(\theta_{enc}, \theta_{dec}) = \sum_{(x,i) \epsilon T} || y-x ||
\end{equation}

The preparatory stage is designed to initialize the model and bring it to a stable state. During this stage, since the model is fed data from the same speaker and trained to make the input and output as similar as possible, it does not effectively learn how to eliminate the speaker identity from the source speech. To address this, we proceed to our proposed stage.

\subsubsection{Speech Classifier}

To better guide the proposed constraint factor, we pretrain a speech classifier that predicts the speaker identity from its feature sequence input, rather than classifying based on latent representation. This approach is more efficient because the latent representations are highly abstract and not as concrete as the source speech sequence. This classifier functions similarly to classic audio speaker classification and is trained on X and I, enabling it to distinguish between different speakers as accurately as possible. The loss function for this classifier is formulated below:

\begin{equation}
L_{clf}(\theta_{clf}) = \sum_{(x,i)} - \log P(i|x)
\end{equation}

\subsection{Proposed Stage}

In our proposed approach, rather than simply adding a classifier to adversarially train the content encoder—which may still retain traces of the source speaker identity due to the need to balance between maintaining input-output similarity and purifying the latent representation from a single source input, and can also result in instability with significant fluctuations—we adopt a multi-task, multi-source, and multi-directional learning strategy. 

Specifically, we incorporate a standard optimizing constraint using a different source for the proposed auxiliary self-destructive amendment steps. In this approach, the destructive component is constrained by the flow of Loss B, while the amendment serves as the primary objective.This method facilitates a more effective separation of linguistic content from speaker identity and tends to result in a simpler and more stable training process.

This proposed stage can be conceived as a repetition and iteration of a combination of two mini-stages, in which we essentially do two things.

\subsubsection{Mini-stage 1}

The first task is to continue training and optimizing the content encoder and decoder, exactly as described in Section 2.1.1, to enhance the model's robustness and reconstruction quality.

\subsubsection{Mini-stage 2}

The second part is the core of our proposed method. Here, we take a step back from being overly aggressive in improving the content encoder and decoder, which might lead to overfitting and potential damage to linguistic content due to excessive focus on uniformity between input and output. 

\begin{figure}[h]
    \centering
    \includegraphics[width=1\linewidth]{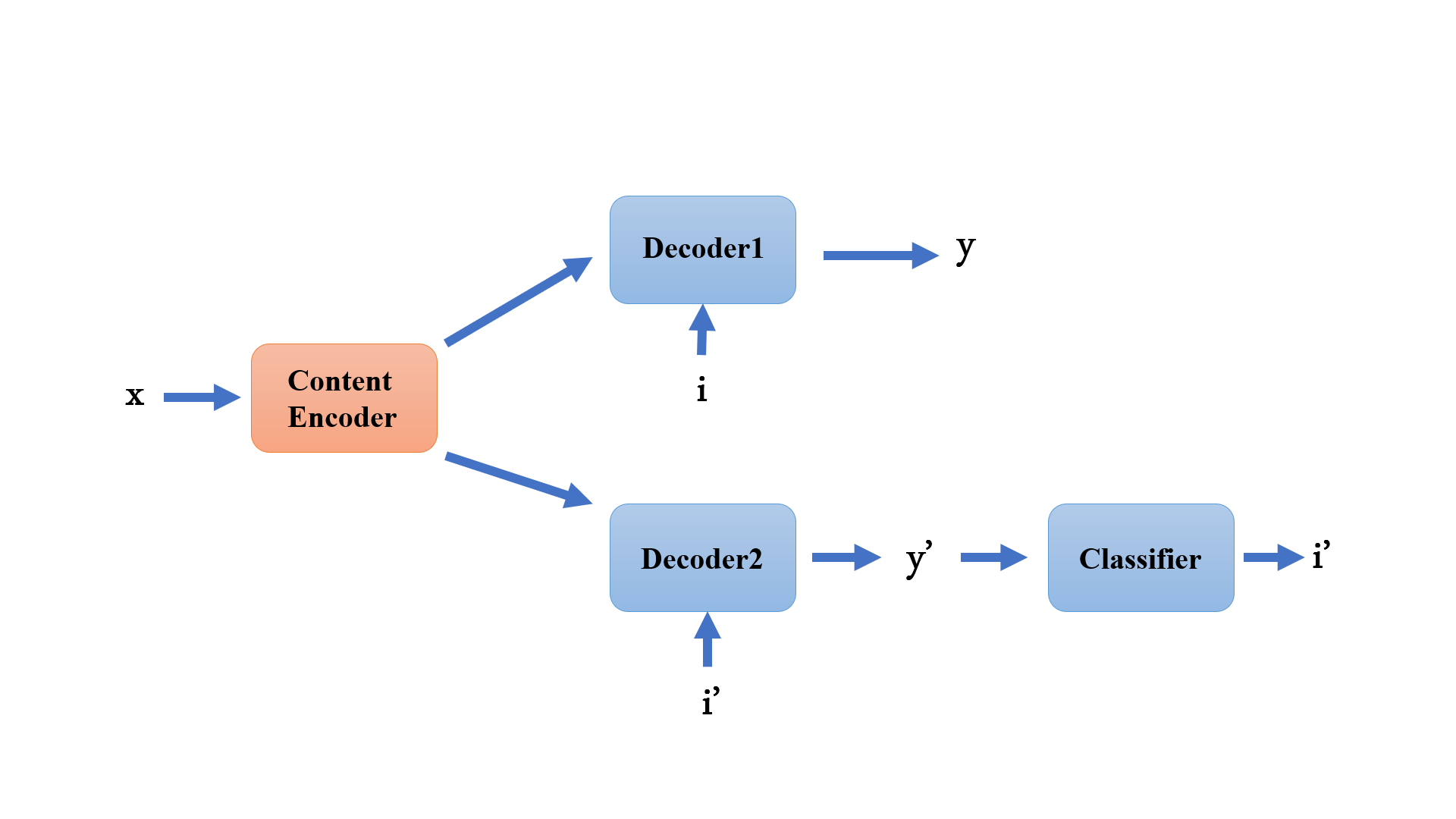}
    \caption{Basic Schematic diagram of mini-stage 2.}
    \label{fig:enter-label}
\end{figure}

As shown in Figure 1, in this second mini-stage, we use two identical decoders with the same parameters, allowing us to input different speaker identities simultaneously during training. First, we use the content encoder to generate latent representations from input x. These latent representations are then fed to two decoder positions. Decoder 1 performs reconstruction (constructing y from x). Meanwhile, Decoder 2 is fed with $i^{\prime}$, a different speaker identity, and aims to generate speech that closely matches the speaker identity $i^{\prime}$ while retaining the linguistic content of x, as examined by the speech classifier.

We add a pretrained classifier to the end of Decoder 2 so it can process $y^{\prime}$ constructed by Decoder 2. The classifier can distinguish between different speakers with consistent parameters, eliminating the need for additional training to adjust to changing latent representations due to the evolving content encoder. This significantly reduces the training time cost. Although the classifier's accuracy may not be very high, it suffices for our purposes, as our focus is on training the content encoder and observing relative improvement.

The loss function for this part is derived from the log probability that $y^{\prime}$ is classified as $i^{\prime}$:

\begin{equation}
L_{low}(\theta_{enc}, \theta_{dec}) = \sum_{(x_i,i^{\prime}) \epsilon T} - \log P({i^{\prime}}|{y^{\prime}})
\end{equation}

During the training process, we aim to minimize this loss.

Additionally, for the upper part, we strive to maximize the distance between x and y when the decoder is fed data from the same speaker using a formula similar to Equation (1). This ensures that the content encoder produces a more speaker-independent representation by leading to a purposeful and accurate self-destructive update.

In this update, we aim to impair the performance of the content encoder and decoder, lowering its ability but not randomly. Instead, this is guided by the Loss B, which will significantly help the system maintain its functionality by making the output $y^{\prime}$ as close as possible to the different source input $i^{\prime}$. Simultaneously, introducing a new domain flow of source-different speaker identity will strengthen the content encoder and decoder, enabling them to generalize better on new, unseen data without adding to the training burden. We observe that by doing so, the quality of both same-speaker and different-speaker inputs is comparable and even enhanced, while the training time cost is reduced to half of what was mentioned in Chou’s paper.

Therefore, the simplest approach is to combine these two losses by addition:

\begin{equation}
L_{back}(\theta_{enc}, \theta_{dec}) = -\lambda L_{upp} + L_{low}
\end{equation}

Where:

\begin{equation}
L_{upp}(\theta_{enc}, \theta_{dec}) = \sum_{(x,i) \epsilon T} || y-x ||
\end{equation}

Here, $\lambda$ is a hyperparameter. We assume the weight for the upper part will be smaller because self-destruction is not the primary goal of the entire stage, but only auxiliary.

This structure forms the combination of the two mini-stages. This combined process will be repeated, thus developing our proposed stage until all the residual information in the latent representation is nearly eliminated, the valuable linguistic content is preserved, and the system operates effectively for the subsequent GAN stage while adapting to speaker-different data. The subsequent GAN stage still follows the design of Chou’s\cite{1}. 

\subsection{Dataset VCTK}

We trained and evaluated our StepBack model using the CSTR VCTK Corpus, which comprises speech data from 110 English speakers with various accents. Each speaker reads approximately 400 sentences selected from newspapers, with each having a unique set of texts. For our experiments, we selected a subset of 20 speakers, evenly split between 10 females and 10 males. The dataset was randomly divided into training and testing sets with a $90\%$ and $10\%$ split, respectively. Data lost during download and transfer were excluded from the analysis, with no adverse impact on the experiment\cite{yamagishi2019vctk}.

This dataset is one of the most widely used in voice conversion, making it easier for us to follow conventions and conduct evaluations. Additionally, the use of non-parallel data, where each speaker reads a unique set of newspaper texts, is particularly relevant to our research objectives. It enables our StepBack model to learn from a diverse range of sentence structures and vocabulary, as well as different genders and nationalities, thereby enhancing its ability to generalize to new, unseen data. This approach ensures that the model is robust and effective in handling various linguistic contexts, contributing to more accurate and versatile speech processing capabilities.

\subsection{Experimental Design}

\subsubsection{Parameter settings and configurations}

The basic design of our model is based on Chou’s model and incorporates the architecture from the CBHG module. The detailed network architecture is outlined in Table 1 and Table 2. To handle variable-length input, we did not use fully-connected layers across time-steps. The convolution-bank is designed to capture local acoustic feature information. We utilized the pixel shuffle layer to generate higher resolution spectra. The term "$emb_{1}(y)$" refers to the speaker embedding in the 1-th layer, acknowledging that the network may require different information at each layer. This embedding is integrated by adding it to the feature map\cite{1}.

All networks employ 1D convolution, except for the discriminator and speaker classifier, which uses 2D convolution to better capture texture. To introduce the necessary noise during training, we applied dropout in the encoder, as recommended. Additionally, we found that adding dropout in the classifier improved the model’s robustness. We used a dropout rate of 0.5 in the encoder and 0.1 in both the speaker classifier and patch discriminator\cite{isola2018imagetoimagetranslationconditionaladversarial}.

Given the challenges in training GANs, we adopted the Wasserstein GAN with gradient penalty (WGAN-GP) objective function to stabilize the training process\cite{12}.

\textbf{Hyper-parameters:} In the initial stage of training, incorporating the distance loss $L_{upp}(\theta_{enc},\theta_{dec})$ (as described in Equation (5)) from the start hindered the autoencoder's ability to accurately reconstruct the spectra. Therefore, we linearly increased the hyper-parameter $\lambda$ from 0 to 0.001 over the first 36,000 iterations to ensure the latent representation gradually became speaker-independent. 

These hyperparameters were selected based on extensive testing and training. Initially, we ensured that every facility was trained for at least the same number of rounds as specified in Chou’s paper. By monitoring the training process using TensorBoard, we were able to determine the optimal hyperparameters and refine the training details. This approach allowed us to balance training costs while enhancing the model's ability to generalize to unseen data.

We used log-magnitude spectrograms as the acoustic features, with spectral analysis and synthesis settings consistent with previous work. The experiments were conducted on a computer equipped with an NVD RTX 3070 GPU.

\begin{table}
    \centering
    \begin{tabular}{|c|c|>{\centering\arraybackslash}p{0.27\linewidth}|}
    \hline
 \textbf{Component}& \textbf{Layer}&\textbf{Details}\\
        \hline
         \multirow{3}*{\textbf{Speaker Classifier}} & Conv block × 5 & C-K-5, stride=2, LReLU, IN\\
         \cline{2-3}
         & Conv layer & C-32-1\\
         \cline{2-3}
         & Output layer & FC-Nspeaker (classifier-1)\\
         \hline
    \end{tabular}
    \caption{VIP model design, highlighting the primary changes. C stands for convolution layer, and FC stands for fully-connected layer. Conv1d-bank-K represents a convolution layer with kernel sizes ranging from 1 to K. LReLU denotes leakyReLU activation, IN indicates instance normalization, Res signifies residual connection, and PS represents the pixel shuffle layer used for upsampling. The kernel sizes K for the discriminator are 64-128-256-512-512.}
    \label{tab:Tabel1}
\end{table}

\begin{table}[h]
    \centering
    \begin{tabular}{|c|c|>{\centering\arraybackslash}p{0.26\linewidth}|} \hline 
         \textbf{Component}&  \textbf{Layer}& \textbf{Details}\\ \hline 
         \multirow{5}*{\textbf{Encoder}}&  Conv-bank block& Conv1d-bank-8, LReLU, IN\\ \cline{2-3} 
         &  Conv block × 3& C-512-5, LreLU
C-512-5, stride=2, LReLU, IN, Res
\\ \cline{2-3} 
         &  Dense block × 4& FC-512, IN, Res\\ \cline{2-3} 
         &  Recurrent layer& Bi-directional GRU-512\\ \cline{2-3} 
         &  Combine layer& Recurrent output + Dense output\\ \hline 
         \multirow{4}*{\textbf{Decoder/Generator}}&  Conv block × 3& embl(y), C-1024-3, LReLU, PS
C-512-3, LReLU, IN, Res
\\ \cline{2-3}  
         &  Dense block × 4& embl(y), FC-512, IN, Res\\ \cline{2-3} 
         &  Recurrent layer& embl(y), Bi-directional GRU-256\\ \cline{2-3}  
         &  Combine layer& Recurrent output + Dense output\\ \hline 
 \multirow{3}*{\textbf{Discriminator}}& Conv block × 5&C-K-5, stride=2, LReLU, IN\\ \cline{2-3}  
 & Conv layer&C-32-1\\ \cline{2-3} 
 & Output layer&Scalar output, FC-Nspeaker (classifier-2)\\ \hline
    \end{tabular}
    \caption{The rest network architectures. C indicates convolution layer. FC indicates fully-connected layer. Conv1d-bank-K indicates con volution layer with kernel size from 1 to K. LReLU indicates leakyReLU activation. IN indicates instance normalization. Res indicates residual connection. PS indicates pixel shuffle layer for upsampling. The kernel size K for discriminator is 64-128-256-512-512.}
    \label{tab:my_label}
\end{table}

\subsubsection{Training and tests phases}

We trained the network using the Adam optimizer with a learning rate of lr = 0.0001, $\beta_1$ = 0.5, and $\beta_2$=0.9. The batch size was set to 32. We randomly sampled 128 frames of the spectrogram with overlap. Initially, we pre-trained the encoder and decoder using the reconstruction loss $L_{pre}(\theta_{enc},\theta_{dec})$ (as described in Equation (1)) for 8000 mini-batches to stabilize the model. Subsequently, we pre-trained the speaker classifier using the classification loss $L_{clf}(\theta_{clf})$ (as described in equation (2)) for 40,000 mini-batches, completing the preparatory stage.

In the proposed stage, we alternated training, performing 4 mini-batches for the encoder and decoder with the loss $L_{pre}(\theta_{enc},\theta_{dec})$  (as described in equation (1)) from mini-stage 1, followed by 1 mini-batch for the same components with the loss $L_{des}(\theta_{enc}, \theta_{dec})$  (as described in equation (4)) from mini-stage 2. This cycle was repeated for a total of 40,000 iterations. Finally, we adopted stage 2 from Chou’s model, where the patch discriminator/generator underwent an additional 50,000 mini-batches of training. !In this stage, the discriminator was trained for 5 iterations for every 1 iteration of the generator.

Overall, our model required only $70\%$ of the training cost compared to Chou’s model, with approximately 548,000 mini-batches for our model versus 808,000 mini-batches in total for Chou’s model.

\subsubsection{Objective Evaluation}

Objective evaluation will involve generating heatmaps of the spectrogram and calculating the global variance for each frequency index of the spectrogram for four conversion examples: M2M (male-to-male), M2F (male-to-female), F2M (female-to-male), and F2F (female-to-female)\cite{10.1121/10.0024876}.

\subsubsection{Subjective Evaluation}

Subjective evaluation will consist of three parts within each section\cite{2}:

\begin{enumerate}
\item Naturalness Comparison: Participants will compare speech samples from Chou’s model and our model, selecting the one that sounds more natural in terms of pitch, tone, stress, and human-likeness.
\item Linguistic Content Preservation: Participants will listen to the source speech and choose the sample that better preserves the linguistic content, including intonation, stress, pitch, and words, without focusing on speaker identity.
\item Speaker Identity Similarity: Participants will listen to the target speech and select the sample from the previous comparisons that is more similar to the target, primarily considering speaker identity.
\end{enumerate}
A total of 20 sections will be evaluated, involving 20 subjects. Preliminary results indicate that our model preserves more linguistic content, maintaining consistent tone, stress, and words, while also achieving comparable or occasionally superior speaker similarity. Notably, our model was trained with fewer mini-batches, indicating lower training costs.

\section{General Discussion}
\subsection{Interpretation of the results}
On one hand, we can view the entire process as using the maximization of differences as an auxiliary task to assist the primary task of the lower part, which aims to improve the content encoder's performance. 

On the other hand, we are engaging in a self-destructive behavior. In the preparatory stage, the content encoder is overly trained to convert speech from the same speaker, resulting in residual traces of the source speech. Therefore, we hypothesize that by making the content encoder less effective when it encounters data from the same speaker (i.e., by increasing the loss), we can help eliminate these residual traces, leading to a cleaner extraction of linguistic content. However, it is crucial to add a constraint to prevent the encoder from becoming completely dysfunctional. This is where the lower part comes in, maintaining solid quality and preventing a total breakdown of the system. 

The combination of these two strategies ensures a balanced approach to refining the content encoder.

\subsection{Strengths and limitations}
Our model, trained with significantly less data, not only delivers comparable or even superior quality in voice conversion but also preserves more linguistic content.

However, due to the nature and purpose of this ablation experiment, we have only modified the major components of the model. In the future, we plan to incorporate more advanced training strategies and architectures that are becoming prominent, such as the renowned ECAPA-TDNN\cite{7}. This model uses statistical pooling to transform variable-length utterances into fixed-length embeddings that effectively characterize speakers.

\section{Conclusion}
Based on extensive experiments and result analysis, the Stepback network significantly improves upon the original classifier-incorporation network in terms of disentanglement completion, linguistic content preservation, and cost reduction. We are eager to see this straightforward strategy applied to new deep learning applications, aiding in feature disentanglement through the addition of constraints to Stepback learning, such as self-destructive amendments.

\bibliographystyle{IEEEtran}
\bibliography{main}

\begin{thebibliography}{10}
\providecommand{\url}[1]{#1}
\csname url@samestyle\endcsname
\providecommand{\newblock}{\relax}
\providecommand{\bibinfo}[2]{#2}
\providecommand{\BIBentrySTDinterwordspacing}{\spaceskip=0pt\relax}
\providecommand{\BIBentryALTinterwordstretchfactor}{4}
\providecommand{\BIBentryALTinterwordspacing}{\spaceskip=\fontdimen2\font plus
\BIBentryALTinterwordstretchfactor\fontdimen3\font minus \fontdimen4\font\relax}
\providecommand{\BIBforeignlanguage}[2]{{%
\expandafter\ifx\csname l@#1\endcsname\relax
\typeout{** WARNING: IEEEtran.bst: No hyphenation pattern has been}%
\typeout{** loaded for the language `#1'. Using the pattern for}%
\typeout{** the default language instead.}%
\else
\language=\csname l@#1\endcsname
\fi
#2}}
\providecommand{\BIBdecl}{\relax}
\BIBdecl

\bibitem{17}
\BIBentryALTinterwordspacing
B.~Sisman, J.~Yamagishi, S.~King, and H.~Li, ``An overview of voice conversion and its challenges: From statistical modeling to deep learning,'' 2020. [Online]. Available: \url{https://arxiv.org/abs/2008.03648}
\BIBentrySTDinterwordspacing

\bibitem{8}
\BIBentryALTinterwordspacing
K.~Qian, Y.~Zhang, S.~Chang, D.~Cox, and M.~Hasegawa-Johnson, ``Unsupervised speech decomposition via triple information bottleneck,'' 2021. [Online]. Available: \url{https://arxiv.org/abs/2004.11284}
\BIBentrySTDinterwordspacing

\bibitem{5}
\BIBentryALTinterwordspacing
T.~Kaneko and H.~Kameoka, ``Parallel-data-free voice conversion using cycle-consistent adversarial networks,'' 2017. [Online]. Available: \url{https://arxiv.org/abs/1711.11293}
\BIBentrySTDinterwordspacing

\bibitem{6}
\BIBentryALTinterwordspacing
T.~Kaneko, H.~Kameoka, K.~Tanaka, and N.~Hojo, ``Cyclegan-vc2: Improved cyclegan-based non-parallel voice conversion,'' 2019. [Online]. Available: \url{https://arxiv.org/abs/1904.04631}
\BIBentrySTDinterwordspacing

\bibitem{9}
\BIBentryALTinterwordspacing
C.-C. Hsu, H.-T. Hwang, Y.-C. Wu, Y.~Tsao, and H.-M. Wang, ``Voice conversion from non-parallel corpora using variational auto-encoder,'' 2016. [Online]. Available: \url{https://arxiv.org/abs/1610.04019}
\BIBentrySTDinterwordspacing

\bibitem{18}
\BIBentryALTinterwordspacing
J.~chieh Chou, C.~chieh Yeh, and H.~yi~Lee, ``One-shot voice conversion by separating speaker and content representations with instance normalization,'' 2019. [Online]. Available: \url{https://arxiv.org/abs/1904.05742}
\BIBentrySTDinterwordspacing

\bibitem{4}
\BIBentryALTinterwordspacing
K.~Qian, Y.~Zhang, S.~Chang, X.~Yang, and M.~Hasegawa-Johnson, ``Autovc: Zero-shot voice style transfer with only autoencoder loss,'' 2019. [Online]. Available: \url{https://arxiv.org/abs/1905.05879}
\BIBentrySTDinterwordspacing

\bibitem{10}
\BIBentryALTinterwordspacing
J.~Serrà, S.~Pascual, and C.~Segura, ``Blow: a single-scale hyperconditioned flow for non-parallel raw-audio voice conversion,'' 2019. [Online]. Available: \url{https://arxiv.org/abs/1906.00794}
\BIBentrySTDinterwordspacing

\bibitem{11}
\BIBentryALTinterwordspacing
V.~Popov, I.~Vovk, V.~Gogoryan, T.~Sadekova, M.~Kudinov, and J.~Wei, ``Diffusion-based voice conversion with fast maximum likelihood sampling scheme,'' 2022. [Online]. Available: \url{https://arxiv.org/abs/2109.13821}
\BIBentrySTDinterwordspacing

\bibitem{yang2024streamvcrealtimelowlatencyvoice}
\BIBentryALTinterwordspacing
Y.~Yang, Y.~Kartynnik, Y.~Li, J.~Tang, X.~Li, G.~Sung, and M.~Grundmann, ``Streamvc: Real-time low-latency voice conversion,'' 2024. [Online]. Available: \url{https://arxiv.org/abs/2401.03078}
\BIBentrySTDinterwordspacing

\bibitem{13}
\BIBentryALTinterwordspacing
S.~Ruder, ``An overview of multi-task learning in deep neural networks,'' 2017. [Online]. Available: \url{https://arxiv.org/abs/1706.05098}
\BIBentrySTDinterwordspacing

\bibitem{kendall2018multitasklearningusinguncertainty}
\BIBentryALTinterwordspacing
A.~Kendall, Y.~Gal, and R.~Cipolla, ``Multi-task learning using uncertainty to weigh losses for scene geometry and semantics,'' 2018. [Online]. Available: \url{https://arxiv.org/abs/1705.07115}
\BIBentrySTDinterwordspacing

\bibitem{16}
\BIBentryALTinterwordspacing
M.~Zhang, X.~Wang, F.~Fang, H.~Li, and J.~Yamagishi, ``Joint training framework for text-to-speech and voice conversion using multi-source tacotron and wavenet,'' 2019. [Online]. Available: \url{https://arxiv.org/abs/1903.12389}
\BIBentrySTDinterwordspacing

\bibitem{3}
\BIBentryALTinterwordspacing
H.~Kameoka, T.~Kaneko, K.~Tanaka, and N.~Hojo, ``Stargan-vc: Non-parallel many-to-many voice conversion with star generative adversarial networks,'' 2018. [Online]. Available: \url{https://arxiv.org/abs/1806.02169}
\BIBentrySTDinterwordspacing

\bibitem{1}
\BIBentryALTinterwordspacing
J.~chieh Chou, C.~chieh Yeh, H.~yi~Lee, and L.~shan Lee, ``Multi-target voice conversion without parallel data by adversarially learning disentangled audio representations,'' 2018. [Online]. Available: \url{https://arxiv.org/abs/1804.02812}
\BIBentrySTDinterwordspacing

\bibitem{yamagishi2019vctk}
\BIBentryALTinterwordspacing
J.~Yamagishi, C.~Veaux, and K.~MacDonald, ``Cstr vctk corpus: English multi-speaker corpus for cstr voice cloning toolkit (version 0.92),'' 2019, [sound]. [Online]. Available: \url{https://doi.org/10.7488/ds/2645}
\BIBentrySTDinterwordspacing

\bibitem{isola2018imagetoimagetranslationconditionaladversarial}
\BIBentryALTinterwordspacing
P.~Isola, J.-Y. Zhu, T.~Zhou, and A.~A. Efros, ``Image-to-image translation with conditional adversarial networks,'' 2018. [Online]. Available: \url{https://arxiv.org/abs/1611.07004}
\BIBentrySTDinterwordspacing

\bibitem{12}
\BIBentryALTinterwordspacing
I.~Gulrajani, F.~Ahmed, M.~Arjovsky, V.~Dumoulin, and A.~Courville, ``Improved training of wasserstein gans,'' 2017. [Online]. Available: \url{https://arxiv.org/abs/1704.00028}
\BIBentrySTDinterwordspacing

\bibitem{10.1121/10.0024876}
\BIBentryALTinterwordspacing
C.~Graham and N.~Roll, ``Evaluating openai's whisper asr: Performance analysis across diverse accents and speaker traits,'' \emph{JASA Express Letters}, vol.~4, no.~2, p. 025206, 02 2024. [Online]. Available: \url{https://doi.org/10.1121/10.0024876}
\BIBentrySTDinterwordspacing

\bibitem{2}
C.~Graham and F.~Nolan, ``Articulation rate as a metric in spoken language assessment,'' in \emph{Interspeech 2019}, 2019, pp. 3564--3568.

\bibitem{7}
B.~Desplanques, J.~Thienpondt, and K.~Demuynck, ``Ecapa-tdnn: Emphasized channel attention, propagation and aggregation in tdnn based speaker verification,'' in \emph{Interspeech 2020}, 2020, pp. 3830--3834.

\end{thebibliography}

\end{document}